\documentstyle[12pt,epsfig,rotating]{article}
\begin{document}
\baselineskip=18pt

\hspace{0.3in}
 
\vspace{0.15in}

\centerline{\bf Studying the scale and $q^2$ dependence of
$K^+\rightarrow\pi^+e^+e^-$ decay}
 
\vspace{0.15in}

\renewcommand{\thefootnote}{\fnsymbol{footnote}}
\centerline{H. Burkhardt\footnote{Also at Shell Centre, School of
Education, University, Nottingham NG8 1BB, England} and J. Lowe\footnote{Also 
at Physics Department, University, Birmingham B15 2TT, England}}

\centerline{\it Physics Department, University of New Mexico, Albuquerque,
NM 87131, USA}
 
\centerline{G. Eilam and M.D Scadron\footnote{Permanent address: 
Physics Department,
University of Arizona, Tucson, AZ 85721, USA}}

\centerline{\it Physics Department, Technion, 32000 Haifa, Israel}
 
\renewcommand{\thefootnote}{\arabic{footnote}}
 
\vspace{0.5in}
 
\begin{abstract}
\noindent We extract the $K^+\rightarrow\pi^+e^+e^-$ amplitude scale at
$q^2=0$ from the recent Brookhaven E865 high-statistics data. We find
that the $q^2=0$ scale is fitted in excellent
agreement with the theoretical
long-distance amplitude. Lastly, we find that the observed $q^2$ 
shape is explained by the combined effect of the pion and kaon form-factor 
vector-meson-dominance $\rho$, $\omega$ and $\phi$ 
poles, and a charged pion loop
coupled to a virtual photon$\rightarrow e^+e^-$ transition.
 
\vspace{0.15in}
 
\end{abstract}
  
\vspace{0.35in} 
 

\centerline{\bf I. INTRODUCTION}
 
\vspace{0.05in}
 
\noindent The decay $K^+\rightarrow\pi^+e^+e^-$ has been studied theoretically
for many years. There is general agreement that the process is
dominated by the long-distance ``bremsstrahlung" graphs, in which a
virtual photon is radiated by the charged kaon or the pion, away from the 
strangeness-changing weak vertex. However, qualitative agreement with
the available experimental evidence on the branching ratio and the form 
factor (i.e. the $q^2$ dependence) has not previously been achieved.
Here, the virtual photon mass, $q^2$, is given by $(p_++p_-)^2$, where 
$p_{\pm}$ are the four-momenta of the $e^{\pm}$. 
Also, most calculations have several unknown parameters, making firm 
predictions difficult. Experimentally, the situation has been dramatically 
improved by the recent availability, from Brookhaven experiment E865 
\cite{865}, of accurate data on both the branching ratio and the $q^2$ 
dependence. In this paper, we fit these data to about 5\% in a calculation 
based on a specific model with no free parameters (except
for relative signs), which fits other 
related data.
 
\vspace{0.05in}
 
\centerline{\bf II. AMPLITUDE AT $q^2=0$}
 
\vspace{0.05in}
 
\noindent The E865 experiment \cite{865} at Brookhaven had measured a new value
of the branching ratio for $K^+\rightarrow\pi^+e^+e^-$. The value depends
on the form of the extrapolation to $q^2=0$. We choose to extrapolate 
with a form factor quadratic in $q^2$, which gives  
$BR(K^+\rightarrow\pi^+e^+e^-)=(2.99\pm 0.06)\times 10^{-7}$
(see Table 1, column 2 of Ref. \cite{865}). From this, one extracts the 
invariant amplitude defined by \cite{ES}
 
\begin{eqnarray}
{\cal M}_{K \to \pi ee}=
A(q^2)(p_K+p_{\pi})^{\mu}\bar{u}_e\gamma_{\mu}v_{\bar{e}}
\end{eqnarray}
 
\noindent giving

\begin{eqnarray}
|A^{\rm exp}(0)|=(4.00\pm 0.18)\times 10^{-9} {\rm GeV}^{-2}\ ,~~
\label{Aexp}
\end{eqnarray}
  
\noindent found from the 3-body phase-space integral for the rate
 
\begin{eqnarray}
\Gamma(K^+\rightarrow\pi^+e^+e^-)=0.2020~
\frac{m_K^5\mid A(0)\mid^2}{3(4\pi)^3}=(1.59\pm0.03)\times 
10^{-23}{\rm GeV}~.
\end{eqnarray}

\noindent As shown in \cite{865}, a quadratic fit 	
provides a better agreement than a linear one for the
same data. 
We will return to the $q^2$ dependence later.
 
\vspace{-2.8in}
\begin{figure}[h]
\hspace{-0.5in}
\epsfig{file=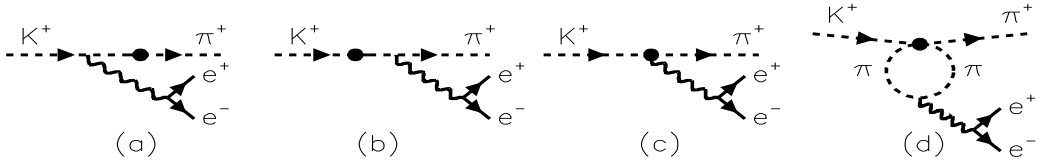,
height=16.0cm,width=10.0cm}
\vspace{-2.2in}
\caption{{\it Graphs for $K^+\rightarrow\pi^+e^+e^-$ through 
a virtual photon. 
(a) and (b) are long-distance graphs, (c) is a short-distance graph and 
(d) is a pion loop term. In each graph, the blob denotes the weak 
(strangeness-changing) vertex and the wavy line is an off-shell   
photon.}} 
\end{figure}
 
\vspace{0.2in}
 
The long-distance (LD) chiral low-energy model, used to calculate the 
bremsstrahlung graphs of Fig. 1(a) and (b), predicts \cite{ES}
 
\begin{eqnarray}
\mid A_{\rm LD}\mid~=~e^2\left|\frac{\langle\pi^+\mid H_W\mid K^+\rangle}
{m_{K^+}^2-m_{\pi^+}^2}\right|~\left|\frac{F_{\pi^+}(q^2)-F_{K^+}(q^2)}
{q^2}\right|~, \label{ALD} 
\end{eqnarray}   

\noindent where $F_{\pi^+}(q^2)$ and $F_{K^+}(q^2)$ are the pion and kaon 
electromagnetic form factors. At $q^2=0$, Eq.~(\ref{ALD}) becomes
 
\begin{eqnarray}
\mid A_{\rm LD}(0)\mid~=~e^2\left|\frac{\langle\pi^+\mid H_W\mid K^+\rangle}
{m_{K^+}^2-m_{\pi^+}^2}\right|~\left|\frac{dF_{\pi^+}}{dq^2}-
\frac{dF_{K^+}}{dq^2}
\right|_{q^2=0}\ .  \label{A} 
\end{eqnarray}  

\vspace{0.05in}

\noindent 
We now present the evaluation of the 
matrix element in Eq.~(\ref{A}), describing how we
obtain $3.9\times 10^{-9}~{\rm GeV}^{-2}$ for $A_{\rm LD}(0)$,
reasonably close to the experimental result in Eq.~(\ref{Aexp}).
The matrix element $\mid\langle\pi^+\mid H_W\mid K^+\rangle\mid$ 
is well established; in Ref. \cite{inter} its value was deduced theoretically
and was confirmed by comparison with
values from ten measured kaon decays, $K_S\rightarrow 2\pi^0$, 
$K\rightarrow 3\pi$, $K_{L,S}\rightarrow 2\gamma$ and 
$K_L\rightarrow\pi^0 2\gamma$, all of which are consistent with
$\mid\langle\pi^+\mid H_W\mid K^+\rangle\mid$ = $\mid\langle\pi^0\mid
H_W\mid K_L\rangle\mid\approx 3.5\times 10^{-8}{\rm GeV}^2$.
Specifically the $K_S \to 2\pi^0$ rate $\Gamma$ gives \cite{PDG}

\begin{equation} 
\mid\langle 2\pi^0\mid H_W\mid K_S\rangle\mid=m_K\sqrt{16\pi\Gamma/p}=
(37.1\pm 0.2)\times 10^{-8}{\rm GeV}~, \label{HW}
\end{equation}
 
\noindent where $p$ is the three-momentum of a $\pi^0$ in the $K_S$ rest
frame. Using partially conserved axial currents (PCAC), this predicts
for the pion decay constant $f_\pi \approx 93$ MeV the
LD scale

\begin{equation}
\mid\langle\pi^+\mid H_W\mid K^+\rangle\mid~\approx~\mid f_{\pi}
\langle2\pi^0\mid H_W\mid K_S\rangle\mid~\approx~ 
3.5\times 10^{-8}{\rm GeV}^2~.  \label{HWK+}
\end{equation} 

For $F_{\pi^+}(q^2)$, there are many experimental measurements
\cite{sjnp18p53} for both positive and negative $q^2$, including our
region $0<q^2<0.125{\rm GeV}^2$. The experimental data close to the 
region of relevance here are shown in Fig. 2. As expected from Vector
Meson Dominance (VMD), these data are well described by a $\rho$ pole; 
the curve in Fig. 2 is the $\rho$-pole, {\it i.e.} VMD prediction, 

\begin{equation} 
F_{\pi^+}(q^2)=\left(1-\frac{q^2}{m_{\rho}^2}\right)^{-1}\Rightarrow 
\frac{dF_{\pi^+}}{dq^2}\left.\right|_{q^2=0}=\frac{1}{m^2_{\rho^0}}
=1.69~{\rm GeV}^{-2}~. \label{Fpiq^2}
\end{equation}
 
\noindent 
The pion charge radius $r_\pi\equiv\sqrt{<r^2_\pi>}$, where
$r^2=6dF(q^2)/dq^2\left.\right|_{q^2=0}$, which in VMD equals
$0.628$~fm, is in agreement with the experimental value \cite{sjnp18p53} 
$r_\pi=(0.63\pm 0.01)$~fm.
Since both $F_{\pi}(q^2)$ and $r_{\pi}$
agree well with the data, the $\rho$-pole 
expression can either be
regarded as an input to our calculation or as an interpolating
function between the experimental points.
Note that up to a small G--parity violation, $\omega$ and
$\phi$ do not contribute to $F_{\pi^+}$.

\vspace{0.2in}
\begin{figure}[h]
\hspace{1.0in}
\epsfig{file=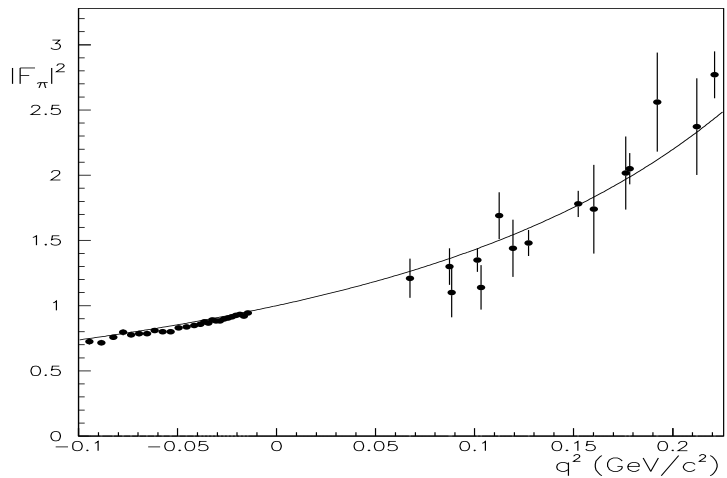,height=8.0cm,width=6.0cm}
\vspace{-0.8in}
\caption{{\it Pion electromagnetic form-factor data \protect\cite{sjnp18p53} 
and the prediction of the $\rho$ pole from VMD, Eq.~(\ref{Fpiq^2}).
The $\rho$ width is included here; it has practically no effect in the
range of $q^2$ relevant for $K^+ \to \pi^+ e^+ e^-$}.}
\end{figure}
 
\vspace{0.05in}
 
For the kaon case, no such comparable data exist, so we need a model 
for $F_{K^+}(q^2)$. In view of the success of the VMD picture for
$F_{\pi^+}(q^2)$, we use the same model for $F_{K^+}(q^2)$, resulting 
in the $\rho$, $\omega$ and $\phi$ pole structure
 
\begin{eqnarray}
F_{K^+}(q^2) & = &
            \frac{N}{2}\left(\frac{g_{\rho ee}}{m_{\rho^0}^2-q^2}+
\frac{g_{\omega ee}}{m_{\omega}^2-q^2}+
\sqrt{2}\frac{g_{\phi ee}}{m_{\phi}^2-q^2}
\right) \Rightarrow \nonumber \\
            &   &\frac{dF_{K^+}}{dq^2}|_{q^2=0}=\frac{n_\rho}{m_{\rho^0}^2}+
                               \frac{n_\omega}{m_{\omega}^2}+
                               \frac{n_\phi}{m_{\phi}^2}=
1.42~{\rm GeV}^{-2}
~, \label{FK+}
\end{eqnarray} 

\noindent where $g_{\rho ee}=5.03$, $g_{\omega ee}=17.06$, $g_{\phi ee}=13.24$ 
(derived from the $e^+e^-$ decay widths), 
and the $\rho^0 K^+ K^-,~\omega K^+ K^-, ~\phi K^+K^-$ $\rm{SU}(3)$
coefficients are $\frac{1}{2}$, $\frac{1}{2}$, $\frac{1}{\sqrt{2}}$,
respectively.
We deduce the $\rho K K$ coupling constant 
from $g_{\rho \pi \pi}$ (obtained using the $\rho$ width)
via SU(3).
We link the normalization $F_{K^+}(0)=1$
to $N=0.037 {\rm GeV}^2$ in Eq.~(\ref{FK+})   
leading to $6 n_{\rho,\omega,\phi}=0.94,3.08,1.99$.
We then find 
that the charged $K$ radius is $r_{K^+}=0.557~{\rm fm}$, in 
very good agreement with the data \cite{rmp50p261}.
The above lead to the prediction $r^2_\pi-r^2_K=
6 \left. \left(dF_{\pi^+}/dq^2-dF_{K^+}/dq^2 \right)\right|_{q^2=0}=
0.063~{\rm fm}^2$,
consistent with the experimental value \cite{plb178p435} 
$(0.100 \pm 0.045)~{\rm fm}^2$.
Substituting into Eq.~(\ref{A}) 
the above values for the derivatives of $F(q^2)$ at $q^2=0$
(see Eqs.~\ref{Fpiq^2} and \ref{FK+}), 
together with Eq.~(\ref{HWK+}) predict

\begin{equation} 
\left|A^{\rm VMD}_{\rm LD}(0)\right|=3.9\times 10^{-9}{\rm GeV}^{-2}\ . 
\label{AVMD0}
\end{equation} 

\noindent Folding in the $\rho$ width into the 
form factors and their derivatives has a negligible effect
for the range of $q^2$ relevant to us here, {\it i.e.} $q^2$
between $0$ and $0.125
{\rm GeV}^2$.

Our assumptions for $F_{\pi^+}(q^2)$ and 
$F_{K^+}(q^2)$ are consistent
with a specific model, which has the VMD and $r_{\pi,K}$ structure built 
in, namely the quark-level linear $\sigma$ model (L$\sigma$M) as discussed 
in Refs. \cite{DS,BRS}. The pion electromagnetic form factor, obtained from a 
$u,\ d$ quark triangle graph for a $\pi^+$ probed by a photon (with 
pseudoscalar $\pi qq$ coupling $g\gamma_5$), predicts 

\begin{equation}
F_{\pi^+}(q^2)=-i4N_cg^2\int_0^1dx\int\frac{d^4p}{(2\pi)^4}\frac{1}
{[p^2-m_q^2+x(1-x)q^2]^2}\ ,  \label{Fpi+} 
\end{equation}
 
\noindent nonperturbatively normalized to $F_{\pi^+}(0)=1$ via the 
quark-loop version of 
the pion decay constant $f_{\pi}$, combined with the quark-level 
Goldberger-Treiman relation $f_{\pi}g=m_q$. To extract the $q^2$
dependence from Eq.~(\ref{Fpi+}) while integrating out the quark momentum
$p$, one can differentiate Eq.~(\ref{Fpi+}), using
$r_{\pi}^2(q^2)=6dF_{\pi^+}(q^2)/dq^2$ for $g^2=(2\pi)^2/N_c$ and
$r_{\pi^+}=1/m_q \approx 0.63$ fm \cite{DS,BRS}:

\begin{equation}   
r_{\pi^+}^2(q^2)=6\int_0^1dx~x(1-x)[m_q^2-x(1-x)q^2]^{-1}~. \label{rpi+}
\end{equation}
 
\noindent Performing the above integration analytically and then making a 
Taylor series expansion 
keeping only the leading terms with $y \equiv q^2/m_q^2$ , one finds
 
\begin{equation}
m_q^2 r_{\pi^+}^2(q^2)=1+\frac{y}{5}+\frac{y^2}{23\ \small 
{\frac{1}{3}}}	
+\frac{y^3}{105}+.\ .\ .~~. \label{rp}
\end{equation}
 
\noindent Since $m_q \approx m_N/3$, so 
$m_{\rho}^2/m_q^2\approx 6$, 
one can approximately express Eq.~(\ref{rp}) in 
VMD language for the form factor 
itself in the low $q^2$ region as in Eq.~(\ref{Fpiq^2}).
Lastly, the coupling $g_{\rho \pi \pi}=6.04$ is taken from
the measured $\rho$ width, with the L$\sigma$M correction 
to VMD predicting \cite
{BRS} $g_{\rho\pi\pi}/g_{\rho ee}=6/5$, very close to the data ratio
$6.04/5.03$. Then from ${\rm SU}(3)$ one gets $g_{\rho K^+K^-}=3.02$.
A test of this approach is to extract $g_{\phi K^+ K^-}=4.58$
from the measured partial width of $\phi \to K^+ K^-$~\cite{PDG},
from which  we find, using ${\rm SU}(3)$ that $g_{\rho K^+ K^-}=3.24$.
This value agrees to better than $10\%$ with its value $3.02$ above. 

In the work of Ref. \cite{ES}, the short-distance (SD) term was 
estimated to cause about a
20\% reduction in the LD term. But more recent work by Dib, Dunietz and 
Gilman \cite{DDG}, based on a much heavier top-quark mass and also 
including QCD corrections, suggests that the SD term is much smaller. 
We therefore neglect the SD term of fig. 1(c) in this work.
  
\vspace{0.05in}
 
Our predicted scale, then, is the LD amplitude of Eq.~(\ref{AVMD0}), 
which agrees 
with the data, Eq.~(\ref{Aexp}), to about 5\%.
 
\vspace{0.05in}
 
\centerline{\bf III. FORM FACTOR}
 
\vspace{0.05in}
 
\noindent To extract the form factor for $K^+\rightarrow\pi^+e^+e^-$ from
Eq.~(\ref{ALD}), we need the $q^2$ dependence of the electromagnetic form
factors, $F_{\pi^+}(q^2)$ and $F_{K^+}(q^2)$. In applying Eq.~(\ref{ALD}),  
we take $F_{\pi^+}(q^2)$ from the data in Fig. 2, parameterised using
the $\rho$ pole (Eq.~(\ref{Fpiq^2})) and for 
$F_{K^+}(q^2)$, we use the equivalent 
expression, Eq.~(\ref{FK+}). We emphasise that we do not rely heavily 
on VMD for $F_{\pi^+}(q^2)$; we use essentially the empirical values. 
The main reason to show the VMD fit to $F_{\pi^+}(q^2)$ is to establish 
that this theory fits the data well, so it is reasonable to use it to 
obtain $F_{K^+}(q^2)$, to apply in Eq.~(\ref{ALD}).
 
\vspace{0.05in}
 
With $F_{\pi^+}(q^2)$, $F_{K^+}(q^2)$ and 
$\langle\pi^+\mid H_W\mid K^+\rangle$ taken as above, the LD part, 
Eq.~(\ref{ALD}), is completely determined. The
resulting form factor, $A_{\rm LD}$, rises with $q^2$, in agreement
with the data, but gives only about $30\%$ of the observed rise in the
form factor, $\mid F\mid^2$, between $q^2=0.03$ and 0.12 ${\rm GeV}^2$.
This is shown by the dashed line in Fig. 3.
 
\vspace{0.5in}
\begin{figure}[h]
\hspace{0.3in}
\epsfig{file=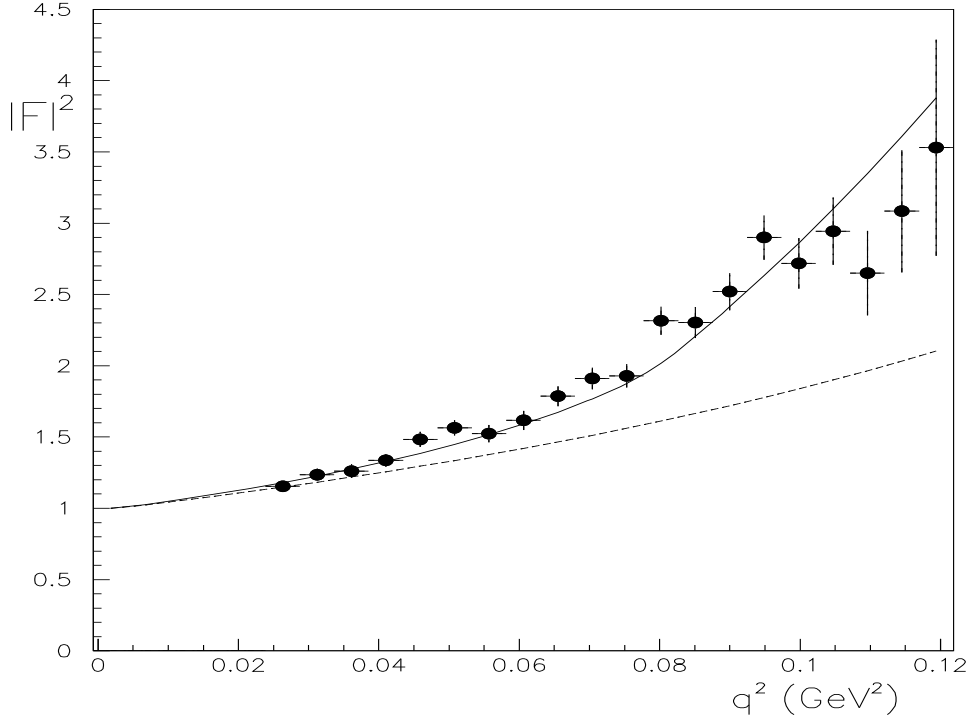,height=12cm,width=8cm}

\vspace{-1.2in}
\caption{{\it Form factor squared, $\mid F\mid^2$, as a function of $q^2$ for 
$K^+\rightarrow\pi^+e^+e^-$. The dashed curve shows $A_{\rm LD}$ and the
solid line shows $A_{\rm LD}+A_{\pi {\rm loop}}$. The black dots are the 
experimental data \protect\cite{865}. The theoretical curves are 
normalized to 1 at $q^2=0$. Complex masses are used here;
inclusion of the imaginary parts of the masses has a small, yet 
visible effect on the high $q^2$ region in this figure}} \end{figure}
\vspace{-0.7cm}
 
Another possible contribution has been discussed by Ecker {\it et al.} 
\cite{EPR} and by D'Ambrosio {\it et al.} \cite{DAM}, namely the 
charged pion loop term, Fig. 1(d). This term is derived in Refs. \cite{EPR} 
and \cite{DAM} using dimensional regularisation. We use here the
expression for $W^{\pi\pi}$ from Ref. \cite{DAM} without any additional 
polynomial. Also,
we do not include terms from Refs. \cite{EPR} and  \cite{DAM} other than 
the loop term; these contributions are calculated explicitly in our 
amplitude $A_{\rm LD}$. Evaluation of the pion loop term requires 
a knowledge of the $K^+\rightarrow\pi^+\pi^+\pi^-$ amplitude. In Refs.
\cite{EPR} and \cite{DAM} this is taken from experiment. Ref. \cite{KS} 
shows that a 
current algebra-PCAC approach for $K^+\rightarrow\pi^+\pi^+\pi^-$ agrees 
with $K_{3\pi}$ data within 5\%, so the $q^2$-dependent part of the 
loop in Refs. \cite{EPR} and  \cite{DAM} is, in fact, compatible with 
the methods 
used here. The relative sign of the pion loop term and the dominant
part of the amplitude, which may be considered as a parameter, was
already established by the experiment \cite{865}.
  
\vspace{0.05in}
 
Adding the amplitude, $A_{\pi {\rm loop}}$, from the pion loop term to
$A_{\rm LD}$ gives the form factor shown as the solid line in
Fig. 3. This agrees quite well with the data. The pion loop term
makes a negligible contribution to the amplitude at $q^2=0$, so it
does not disturb the agreement with experiment of the scale,
$A(0)$, discussed in Sec. II and resulting in Eq.~(\ref{AVMD0}).
  
\vspace{0.05in}
 
\centerline{\bf IV. CONCLUSIONS}
 
\vspace{0.05in}
 
\noindent In summary, we have shown that the recent Brookhaven data for
$K^+\rightarrow\pi^+e^+e^-$ are in qualitative agreement, both in 
the $q^2=0$ scale and the form factor, with a calculation in which 
two processes are included. The experimental amplitude, 
$|A^{\rm exp}(0)|=(4.00\pm 0.18)\times 10^{-9}{\rm GeV^{-2}}$ 
(Eq.~(\ref{Aexp})), is about 
$5\%$ around the simple VMD bremsstrahlung
prediction of $\left|A_{\rm LD}(0)\right|=3.9\times 10^{-9}{\rm GeV^{-2}}$ 
in Eq.~(\ref{AVMD0}), having dropped the SD term according to Ref. 
\cite{DDG}. 

This agreement is as good as can be expected from such a simple model.
In particular folding in a $K^*$ pole through $K^+ \to \pi^+ K^{*0}
\to \pi^+ e^+ e^-$ and a small, about $10\%$  SD contribution 
to the amplitude \cite{Singer}, may modify
our prediction slightly. Note that as more terms are added, one
increases the danger of double counting.
Each of the graphs discussed 
here has been calculated before, though all have never been taken 
together in one calculation. This level of agreement is reasonable in
view of the simplicity and combination of the models used.

There are no free parameters in our treatment,
except for the relative sign between the pion loop and the rest of the 
amplitude and a relative sign between the couplings
in Eq.~(\ref{FK+}); all quantities are taken 
from experiment, or from models that are known to fit other low-energy 
kaon or pion processes well. We do not attempt to estimate theoretical 
errors on our calculation, since these are to some extent dependent on 
one's point of view. For example, the pion electromagnetic form factor, 
shown in Fig. 2, could equally well be regarded as taken from experiment, 
with its associated errors, or from the VMD  prediction which is 
essentially free from error within the model. Similarly, the value of 
$\mid\langle\pi^+\mid H_W\mid K^+\rangle\mid$ is derived from theoretical 
arguments, but is confirmed by the fit to no less than ten other kaon 
decays [2,3]. Either of these sources could be regarded as the origin of 
our value for $\mid\langle\pi^+\mid H_W\mid K^+\rangle\mid$ and hence as 
the source of error in it.

Concerning the recent paper 
\cite{Lichard} on $K^+ \to \pi^+ ee$, we agree with its recommended 
form factor structure (as in our Eq.~(\ref{ALD}); compare with
its Eqs. (2), (7) and (8)). However we disagree about the scale
of Eqs. (7) and (8) in  \cite{Lichard} which seems more akin
to the small $\Delta I=3/2$ $K \to \pi \pi$ tree level transition rather
than to the much larger $K^0 \to \pi \pi$ $\Delta I=1/2$ LD transition, 
as in our Eqs.~(\ref{HW}) and (\ref{HWK+}).

As for Chiral Perturbation Theory (ChPT)
applied to $K^+ \to \pi^+ ee$~\cite{Isidori}, the required ${\cal O}(p^6)$
expression contains about 100 parameters. Moreover, the underlying 
$\Delta I=1/2$ LD scale as in our Eq.~(\ref{HWK+}) disappears for
$N_C \to \infty$ \cite{Isidori}.

\vspace{0.1in}
 
\centerline{\bf ACKNOWLEDGEMENTS}
 
\vspace{0.05in}
 
We are grateful to T. Goldman, G.J. Stephenson Jr., P. Singer and 
F. Krauss for
useful discussions. We acknowledge support from the US DOE.
GE would like to thank the US-Israeli Binational
Science Foundation and the Fund for the Promotion of Research at the
Technion for partial support. MDS appreciates partial support 
from the Technion.


\vspace{0.5in}
 
\vspace{0.1in}

\begin{thebibliography}{99}

\bibitem{865}
E865 collaboration, R. Appel {\it et al}., 
Phys. Rev. Lett. {\bf 83}, 4482 (1999).

\bibitem{ES}
G. Eilam and M.D. Scadron, Phys. Rev. {\bf D31}, 2263
(1985).

\bibitem{inter}
M.D. Scadron and R.E. Karlsen, Intersections 91
Conference, AIP Conference Proceedings no. 243, ed. W.T.H. Van
Oers, 1992, p. 627; Nuovo Cimento A{\bf 106}, 113 (1993).
 
\bibitem{PDG}
Particle Data Group, D.E. Groom {\it et al.}, Eur. Phys. J.
{\bf C15}, 1 (2000).
 
\bibitem{quatordici} R.E. Karlsen and M.D. Scadron, Phys. Rev. 
{\bf D44}, 2192 (1991). 

\bibitem{sjnp18p53}
S.F. Bereshnev {\it et al}., Sov. J. Nucl. Phys. {\bf 18}, 
53 (1974); {\bf 24}, 591 (1976);
A. Quezner {\it et al}., Phys. Lett. {\bf B76}, 512 (1978);
I.B. Vasserman {\it et al}., Sov. J. Nucl. Phys. {\bf 33}, 368 (1981);
 E.B. Dally  {\it et al}., Phys. Rev. Lett. {\bf 48}, 375 (1982); 
A.F. Grashin and W.V. Lepeshkin, Phys. Lett. B{\bf 146}, 11 (1984); 
L.M. Barkov et al., Nucl. Phys. {\bf B256}, 365 (1985);
S.R. Amendolia {\it et al}., Nucl. Phys. {\bf B277}, 168 (1986). The last
of these references contains an excellent summary of all earlier work.
  
\bibitem{rmp50p261}
T.H. Bauer, R.D. Spital, D.R. Yennie and F.M. Pipkin, Rev. Mod. Phys.
{\bf 50}, 261 (1978); Erratum-{\it ibid.} {\bf 51}, 407 (1979);
L. Ametller, C. Ayala and A. Bramon, Phys. Rev. {\bf D24}, 233 (1981);
C. Ayala and A. Bramon, Europhys. Lett. {\bf 4}, 777 (1987).
Eq.~(11) in the latter reference 
has VMD forms similar 				
to our results in Eqs.~(\ref{Fpiq^2}, \ref{FK+}) above.

 
\bibitem{plb178p435}
S.R. Amendolia {\it et al}., Phys. Lett. {\bf B178}, 435 (1986).

\bibitem{DS}
R. Delbourgo and M.D. Scadron, Mod. Phys. Lett. {\bf A10}, 251(1995);
T. Hakioglu and M.D. Scadron, Phy. Rev. {\bf D43}, 2439 (1991).

\bibitem{BRS}
A. Bramon, Riazuddin and M.D. Scadron, J. Phys. G {\bf 24},
1 (1998); hep-ph/9709274; also see N. Paver and M.D. Scadron, Nuovo
Cimento A{\bf 78}, 159 (1983); 
The equivalent result $r^2_{\pi}=N_c/4\pi^2f^2_{\pi}$ was earlier
obtained by R. Tarrach, Z. Phys. C{\bf 2}, 221 (1979); S.B. Gerasimov,
Sov. J. Nucl. Phys. {\bf 29}, 259 (1979); V. Bernard, B. Hillar and
W. Weise, Phys. Lett. B{\bf 205}, 16 (1988).

\bibitem{DDG}
C.O. Dib, I. Dunietz and F.J. Gilman, Phys. Rev. D{\bf 39}, 2639 (1989), 
based on the formalism of F.J. Gilman and M. Wise, Phys. Rev. D{\bf 21}, 
3150 (1980); See also L. Bergstr\"om and P. Singer, Phys. Rev. D{\bf 43}, 
1568 (1991).

\bibitem{EPR}
G. Ecker, A. Pich and E. deRafael, Nucl. Phys. {\bf B291},
692 (1987).

\bibitem{DAM}
G. D'Ambrosio, G. Ecker, G. Isidori and J. Portol\'es,
JHEP, {\bf 8}, 4 (1998); G. Ecker, private communication.
 
\bibitem{KS}
R.E. Karlsen and M.D. Scadron, Phys. Rev. D{\bf 45},
4108 (1992).

\bibitem{Singer}
See L. Bergstr\"om and P. Singer in \cite{DDG} and
P. Singer, hep-ph/9607429.

\bibitem{Lichard}
P. Lichard, Phys. Rev. D{\bf 61}, 073010 (2000).

\bibitem{Isidori} For a summary see: G. Isidori, hep-ph/0011017.
\end{thebibliography}
\end{document}